# Effect of the reduction process on the field emission performance of reduced graphene oxide cathodes

**Labrini Sygellou[1*], George Viskadouros[3,4], Costas Petridis[3], Emmanuel Kymakis[3], Costas Galiotis[1,5], Dimitrios Tasis[1,6], Emmanuel Stratakis[1,7*]**

[1]Foundation of Research and Technology Hellas, Institute of Chemical Engineering and High Temperature Chemical Processes (FORTH/ICE-HT), P.O. Box 1414, Gr-26504, RionPatras, Greece

[2]Foundation of Research and Technology Hellas, Institute of Electronic Structures and Laser (FORTH/IESL), Gr-70013, Heraklion Crete, Greece

[3]Center of Materials Technology and Photonics & Electrical Engineering Department Technological Educational Institute (TEI) of Crete, Heraklion 71004 Crete, Greece

[4]Department of Mineral Resources Engineering, Technical University of Crete, Chania, 731 00, Crete, Greece

[5]Department of Chemical Engineering, University of Patras, 26504 Patras, Greece

[6]Department of Chemistry, University of Ioannina, 45110 Ioannina, Greece

[7]Departmentof Materials Science and Technology, University of Crete, Heraklion, 71003, Greece

**ABSTRACT**

The electron field emission (FE) properties of reduced graphene oxide (rGO) cathodes produced by three different reduction methods were assessed and compared. In particular, chemical reduction techniques, using either NaOH or KOH as reducing agent, were compared with thermal reduction (TR) methods. X-ray photoelectron spectroscopy (XPS) measurements revealed that different reduction techniques led to different GO lattice parameters. Furthermore, the work function measured with ultraviolet photoelectron spectroscopy (UPS) varied among the samples giving rise to different electron emission characteristics. In particular, the cathodes prepared by the TR method presented the best FE performance, showing a turn-on field of as low as ~1.8 V/μm and a field enhancement factor of ~1300, which was very close however to that shown by the NaOH-reduced sheets. The worst FE properties was exhibited by the KOH-reduced nanosheets. In view of the above results, the role of the different reduction techniques as well as the final rGO lattice characteristics with regards to the emission performance are evaluated and discussed.

## 1. INTRODUCTION

Cold cathode FE is the process for which electrons tunnel through the potential barrier of a material's surface subjected to a strong electric field. The morphological and geometrical characteristics of the emitting sites are crucial for this mechanism, that is, the higher the aspect ratio (the ratio of lateral size to thickness) the lower the applied electric field required for the generation of field emission currents. In the past few decades, intense research effort has been devoted to the design and fabrication of cold cathode electron field emitters for applications in vacuum microelectronic devices, such as electron guns, microwave power amplifiers, X-ray tubes [1,2] neutralizers, space propulsion devices [3] and FE-based electronic devices, including flat lamps and flat panel FE displays (FEDs). Since the isolation of graphene, a two-dimensional carbon allotrope, there is an increasing interest for graphene-based electronic devices. Graphene is a material with unique electrical properties that exhibits high mobility of charge carriers [4]. Also, the high aspect ratio of graphene sheets together with the presence of sub-nanometer edges allow the extraction of electrons at low threshold electric fields with high geometric field enhancement. Finally, graphene exhibits excellent mechanical properties, including high mechanical strength and flexibility, as well as high thermal conductivity. All these properties make graphene an excellent candidate for many nanoelectronic applications including high-performance FE devices [5,6,7].

Most of the efforts to date to implement graphene as a FE, using various production methods such as mechanical cleavage [8], chemical vapor deposition [9], and chemical exfoliation [10], have showed promising electron emission properties; however, further work and research effort is required in order to develop large area graphene-based FE devices. Chemical exfoliation of bulk graphite using strong oxidizing agents (Hummers method) is quite useful for the large scale synthesis of dispersable graphene flakes particularly in the form of graphene oxide (GO) [11,12]. GO has a similar single atomic layer structure, but unlike graphene, the presence of oxygen-containing labile groups attached onto $sp^3$-hybridized carbon atoms leads to a significant decrease of its conductivity. This property is undesirable for most of electronic applications and several methods have been proposed in order to reduce GO towards the formation of rGO, and in this way to partially restore its conductivity. Reduction methods include the utilization of chemical reducing agents e.g. hydrazine[13] and zinc powder [14], thermal annealing at high temperature under inert atmosphere [15], and application of either electric [16] or electromagnetic field [17]. Despite of its inferior conductivity, compared to pristine graphene, rGO is readily processable in solution and thus can be deposited in large areas onto any type of substrate, enabling simple and cost-effective fabrication of FE devices for display applications.

In the present contribution, we investigate the effect of the reduction process on the FE performance of rGO nanosheets. For this purpose, we perform a comparative study of the FE performance of rGO-based cathodes, which have been derived by three different GO reduction methods. In particular, cathodes based on material chemically reduced by either aqueous NaOH or KOH were directly compared with rGO derived following TR process. In all cases, rGO layers were deposited via a fast and facile solution-based method onto $n^+$Si conductive substrates. Chemical analysis, using XPS, revealed that the various reduction methods lead to different GO lattice chemistries, giving rise to remarkable differences in the FE characteristics. The FE performance was quantitatively validated via work function (WF) measurements of the respective rGO materials, performed by UPS. It is revealed that the oxygen content within the graphene lattice is the most crucial parameter affecting the FE performance of rGO-based cathodes.

## 2. EXPERIMENTAL SECTION

### 2.1 Preparation of the starting GO solution

The GO material was synthesized by a modified Hummer's method via adopting a two-step oxidation of graphite flakes [18]. Details are given in a previous work of our group [19]. For the preparation of stable aqueous suspensions highly enriched in monolayer GO sheets, a certain quantity of GO was exfoliated by bath sonication for a period of one hour. Aggregated GO sheets were discarded through centrifugation of sonicated suspension at 4000 rpm for 20 min. Thus, individually suspended GO platelets were isolated in the supernatant part, at a concentration of about 1 mg/mL.

### 2.2 Reduction strategies of GO material

*2.2.1 Chemical Reduction using NAOH as reducing agent*

In the case of NaOH-mediated reduction of GO [20], two different batches were prepared; 370 ml of centrifuged GO solution was mixed with sodium hydroxide (NaOH) solution (total $C_{NaOH}$ was ranged between 0.04 and 0.056 M in our tests). The resulting mixture was stirred for 4 hours at room temperature and was filtered through a polycarbonate membrane (400 nm pore size). Sample rGO/NaOH(1) was denoted as the material, which stayed on top of the membrane. The material passed through the polycarbonate membrane (black-colored filtrate) was further filtered through a Teflon (PTFE) membrane with 200 nm pore size. This sample was denoted as rGO/NaOH(2). So, both samples are actually NaOH-treated GO towards rGO.

*2.2.2 Chemical Reduction using KOH as reducing agent*

The sample denoted as rGO/KOH was prepared by suspending GO platelets in an aqueous 1 M potassium hydroxide (KOH) solution and subsequent filtration of the mixture [21]. After washing repeatedly with water, the KOH-based reduced GO material was dried under vacuum.

*2.2.3 Thermal reduction of GO*

The thermally reduced GO sample(rGO/TR) was actually derived by gradually heating a GO film from room temperature up to 200 C and by keeping at that temperature for 1 hour. The heating was done under vacuum conditions. This thermal process resulted in deoxygenation/reduction of GO.

**2.3 XPS-UPS measurements**

The surface analysis studies were performed in a UHV chamber (P<$10^{-9}$ mbar) equipped with a SPECS LHS-10 hemispherical electron analyzer. The XPS measurements were carried out at room temperature using unmonochromatized AlKa radiation under conditions optimized for maximum signal (constant ΔE mode withpass energy of 36 eV giving a full width at half maximum (FWHM) of 0.9 eV for the Au 4f7/2 peak). The analyzed area was an ellipsoid with dimensions 2.5 x 4.5 mm$^2$. The XPS core level spectra were analyzed using a fitting routine, which allows the decomposition of each spectrum into individual mixed Gaussian-Lorentzian components after a Shirley background subtraction. Errors in our quantitative data are found in the range of ~10% (peak areas) while the accuracy for BEs assignments is ~0.1 eV.

The Ultra Violet Photoelectron Spectroscopy studies were performed in a UHV chamber with a SPECS LHS-10 hemispherical electron analyzer. The UPS spectra were obtained using HeI irradiation with hv=21.22eV produced by a UV source (model UVS 10/35). During UPS measurements the analyzer was working at the Constant Retarding Ratio (CRR) mode, with CRR=10. A bias of −12.29 V was applied to the sample in order to avoid interference of the spectrometer threshold in the UPS spectra. The high and low binding energy and highest occupied molecular orbital (HOMO) cutoff positions were assigned by fitting straight lines on the high and low energy cutoffs of the spectra and determining their intersections with the binding energy axis. Regarding measurement errors it should be noted that an error of ±0.1 eV is assigned to the absolute values for ionization energies, work function and other UP-spectra cutoff features where the error margin is significant, due to the process of fitting straight lines.

**2.4 Field Emission Measurements**

The FE cathodes were prepared via drop casting of an initial rGO solution (10 mg/ml) on n$^+$ Si wafers. In particular, the initial rGO aqueous suspension was sonicated for 15min and then drop-casted over a Si substrate. Then the rGO suspension was allowed to get dried overnight, resulting in a uniform coating onto Si. The surface morphology of the film was examined by field emission scanning electron microscopy (FESEM JEOL-JSM7000F) both prior and after FE. Following the FE process, there was no apparent explosive destruction to the film, which can be associated with discharge current phenomenon. FE measurements were performed under high vacuum conditions (< $10^{-6}$ Torr), using the samples as cold cathode emitters in a short - circuit protected planar diode system. Details for the experimental setup can be found elsewhere [22]. Current density – electric field (*J-E*) curves were taken at the distance between anode-cathode was controlled by a stepper motor and found that field emission characteristics are not influenced by the anode height. All measurements presented here were performed at *t*=200 μm. Several emission cycles were taken in order to verify the stability and the reproducibility of the *J-E* curves. A voltage with variable sweep step, supplied by a HV source (PS350-SRS), was applied between the anode and the cathode to extract electrons. The emission current was measured using an auto-arranging digital Pico-ammeter (Keithley 485) protected against high voltage surges by a MOSFET limiter. The stability of the emission current over time was examined by monitoring the evolution of the emitted current density over a long time period of continuous operation.

## 3. RESULTS AND DISCUSSION

Figure 1 shows representative top (1a) and cross-sectional (1b) SEM views, showing that the rGO layers are very smooth and uniformand have a thicknessof the order of 1 μm.

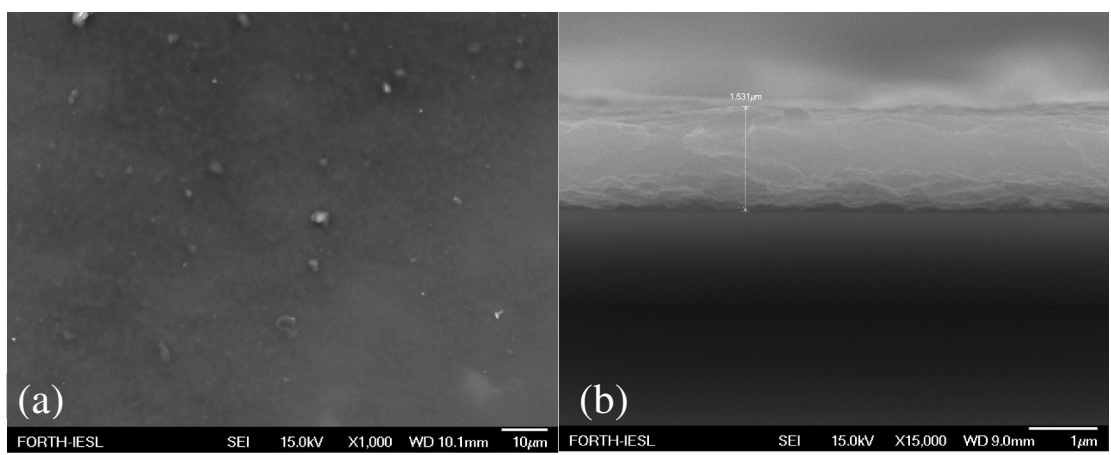

**Figure 1:** Typical top (a) and cross-sectional (b) SEM images of rGO layers.

SEM micrographs indicated that there are no remarkable differences in the surface morphology among the four types of rGO layers prepared. XPS survey scans (not shown) revealed, except of carbon and oxygen, the presence of potassium in rGO/KOH sample and sodium in both rGO/NaOH samples. Figure 2 shows the C1s core level peak of all samples, while for rGO/KOH sample the K2p doublet is shown in the same plot.

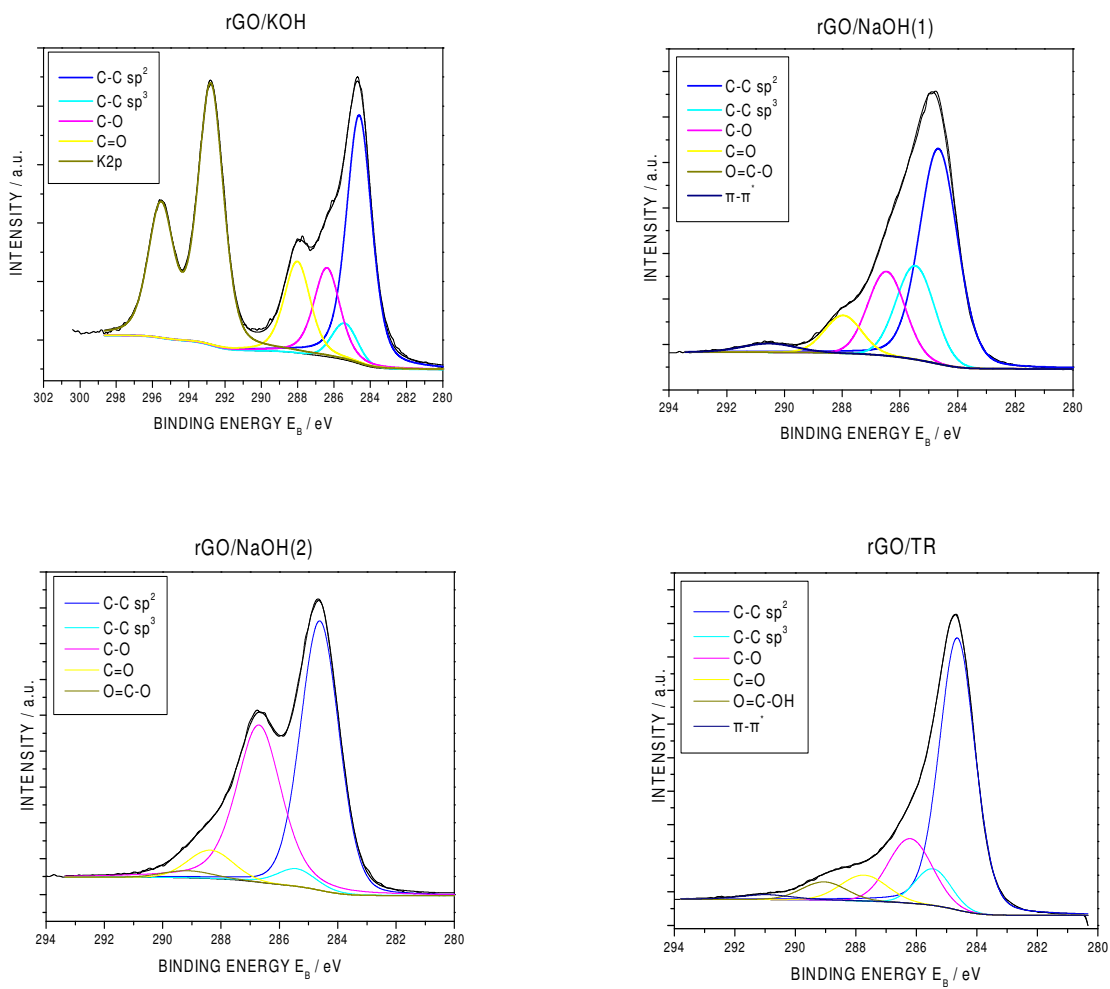

**Figure 2 :** Deconvoluted C1s XPS core level peak of (a) rGO/KOH sample, (b) rGO/NaOH(1), (c) rGO/NaOH(2) and (d) rGO/TR samples. In (a) a combined XP C1s - K2p core level window is shown.

Each C1s peak is analyzed in five components the assignments of which are given elsewhere [23]. Briefly, at 284.7eV assigned to $sp^2$ C-C hybridization, at 285.6eV to $sp^3$ C-C hybridization, at 286.5eV to the C-O hydroxyls and epoxy-ether groups, at 288.1eV to C=O groups, at 289.2eV to the O=C-OH groups and at ~291eV to the π-π* transition peak. The corresponding intensity ratio of carbon-oxygen to carbon-carbon bonds (C-O/C-C), as derived from the deconvolution of C1s peak is shownin table I. It is evident that following reduction for samples rGO/TR and rGO/NaOH(1) there is a significant decrease in the C-OH peaks. For rGO/KOH sample the C=O component has significantly higher intensity in relation to the C-O component than the other samples. In Table I also the % Carbon-Oxygen bonds (C-O, C=O, COOH) and % carbon-carbon bonds ($sp^2$ and $sp^3$) to the total C1s peak intensity, as well as the % oxygen-to-carbon atomic ratio are showed. The $K2p_{3/2}$ peak appeariang at 292.8eV is assigned to $K^+$ ions which are probably bounded with the carboxylic acid groups at the edges of graphene oxide sheets [24]. This may be the reason of the high C=O component in comparison with the other samples.

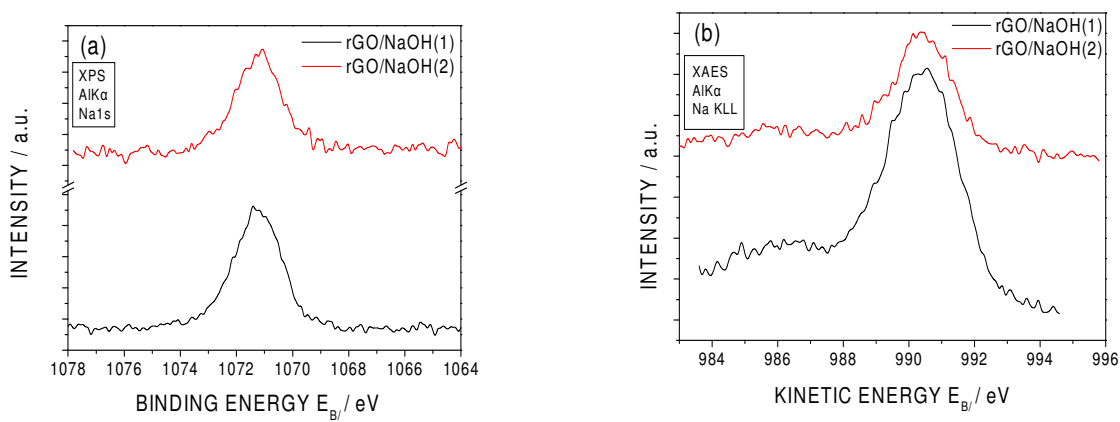

**Figure 3 :** (a) XPS Spectra of Na1s and (b) XAES Spectra of NaKLL of rGO/NaOH(1) and rGO/NaOH(2) samples.

Figure 3 shows the Na1s XPS peak and NaKLL Auger peak. The Binding Energy of Na1s is 1071.2±0.1eV and the kinetic energy of NaKLL is 990.4±0.1eV so the modified Auger parameter is 2061.6±0.1eV and is attributed to Na-C-O bonds, probably COONa [25]. The peak intensity ratio Na1s to C1s peak intensity is 0.35 for rGO/NaOH(1) and 0.19 for rGO/NaOH(2) indicating the higher amount of Na on the rGO/NaOH(1) surface.

The ionization potential (IP: the energy difference between the valence band maximum and the vacuum level) can be determined from the analysis of the respective UPS spectra pesented in Figure 4. The spectra consist of two areas of interest: the region near the Fermi level (low binding energy side, shown on the left in Figure 4) where the valence band maximum relative to the Fermi level can be determined (HOMO cutoff) and the high binding energy cutoff, (right side in Figure 4), where the electrons have barely enough energy to overcome the WF of the material. From the binding energy of this cutoff, the WF of the surface can be determined by subtracting the spectrum's width (i.e. the energy difference between the Fermi level and the high binding energy cutoff), from the HeI excitation energy. The IP is calculated by adding the absolute values of the measured WF and the HOMO cutoff and the results shown in table I. The results shows that the rGO/TR and rGO/NaOH(1) samples have similar WF and IP. This is not the case the rGO/KOH and rGO/NaOH(2) samples. The WF is higher in the case where the oxygen content is more pronounced in agreement with literature [26]. The same trent is in the HOMO cut-off region, where in the rGO/KOH and rGO/NaOH(2) samples there are no states in the Fermi Level in contrast with the rGO/TR and rGO/NaOH(1) samples.

**Table I:** FE properties (turn on field, field enhancement), intensity ratio of carbon-oxygen to carbon-carbon bonds (C-O/C-C as derived from the deconvolution of C1s XPS peak), the % oxygen to carbon atomic ratio (as derived from the intensity ratio of C1s and O1s XPS peaks), Work Function and Ionization Potential (as derived from the UPS analysis) of different rGOfilms. The ± values denote the standard deviation of each measured or estimated quantity.

| Sample | rGO/KOH | rGO/NaOH(1) | rGO/NaOH(2) | rGO/TR |
|---|---|---|---|---|
| Turn-on Field $F_{to}$ (±0.10)(V/μm) | 7.0 | 2.5 | 9.5 | 1.8 |
| Field Enhancement Factor $\beta$ | 199±10 | 1351±60 | 46±5 | 1327±60 |
| C-O/C-C | 1.02 | 0.35 | 0.89 | 0.40 |
| O/C atomic ratio | 0.60 | 0.39 | 0.48 | 0.23 |
| Work Function (±0.10)/ eV | 4.80 | 4.50 | 4.85 | 4.50 |
| Ionization potential(±0.10)/ eV | 5.7 | 4.8 | 5.9 | 4.5 |

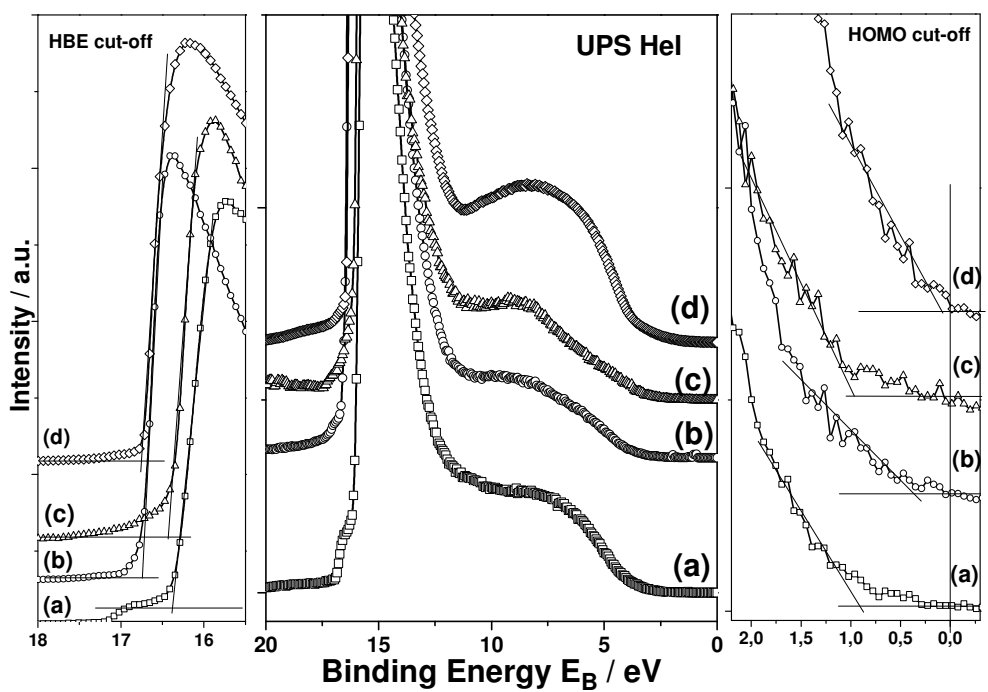

**Figure 4 :** HeI UPS spectra of (a) rGO/KOH sample, (b) rGO/NaOH(1), (c) rGO/NaOH(2) and (d) rGO/TR samples. The full spectra are shown in the center part. On the right part, the region near the spectrometer's Fermi level (low binding energy side, where electrons from the highest molecular orbital ejected) and on the left, the high binding energy cut-off regions are shown magnified for clarity.

Figure 5a shows, on a log-log plot, the current density, *J*, measured as a function of the electric field, *E* for the various rGO layers with the optimum density, obtained by the different reduction methods described in the experimental section. No emission was observed from a cathode prepared using the pristine GO solution. The corresponding turn-on fields, $E_{to}$s are shown in Table 1. $E_{to}$, is defined as the average macroscopic field needed to extract 25 pA/cm$^2$. It is clear that the best emission thresholds are obtained for thermally reduced rGOsheets, while similar performance is measured for the rGO/NaOH(1) sample. On the contrary the rGO/KOH and the rGO/NaOH(2) samples exhibited an inferior performance as compared to the other two samples.

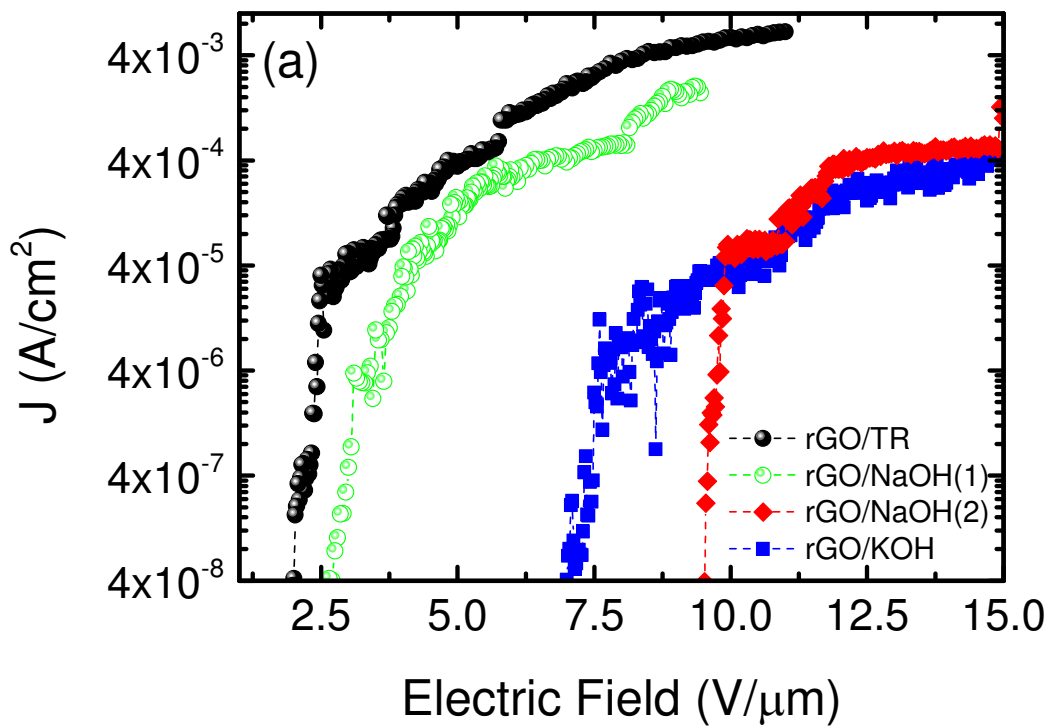

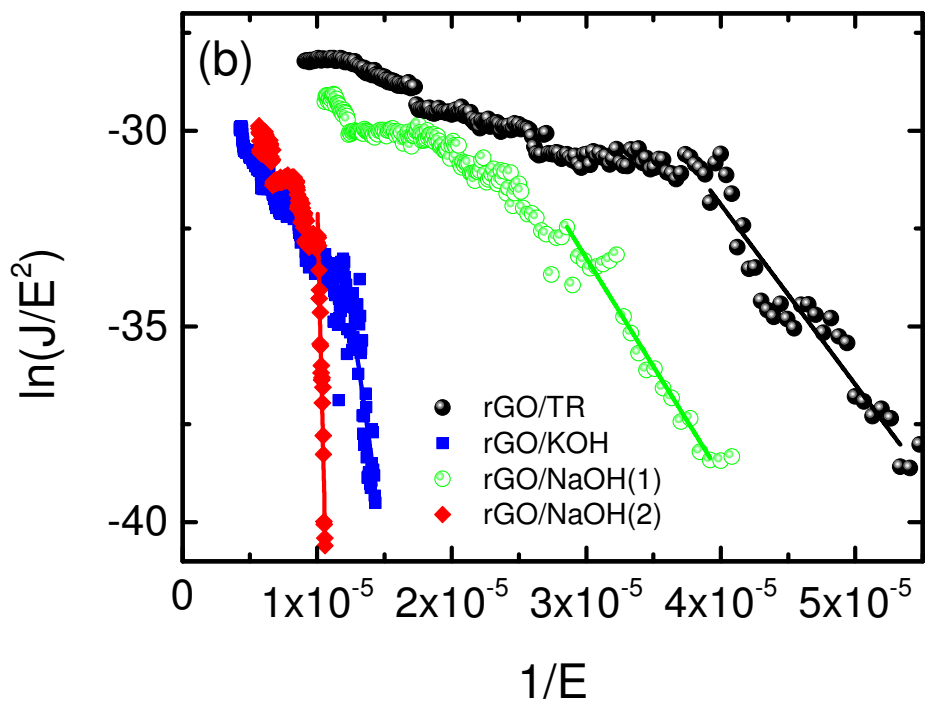

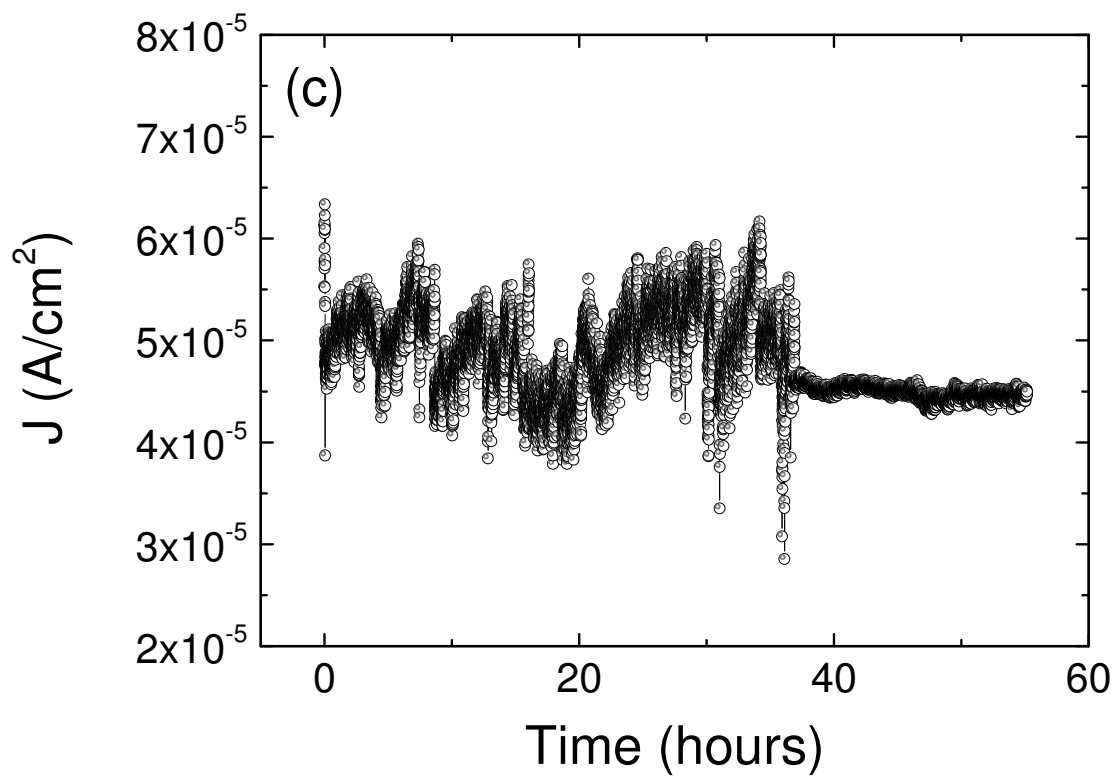

**Figure 5:** (a) Logarithmic plot of current density-field (*J-E*) emission characteristics of rGO films prepared using different reduction methods, shown in the legend and defined in the text; (b) Fowler-Nordheim plots corresponding to the J-E curves of (a); (c) Emission current stability over time at a constant voltage bias, for the best performed rGO/TR cathode.

We analyze our field emission using the classic theory of Fowler-Nordheim (FN), about field-assisted tunneling process in which the current density, $J$, depends on the local microscopic field at the emitter, $E_{loc}$, according to the relationship:

$$J = AE_{loc}^2 \exp\left(-\frac{b_{FN}}{E_{loc}}\right) \quad (1)$$

where $A$ is a constant that depends on the actual emitting surface structure, $b_{FN} = 0.94 \cdot B \cdot \Phi^{3/2}$ with B=6.83x10$^7$ Vcm$^{-1}$eV$^{-3/2}$, and $\Phi$ is the WF of the material in eV. $E_{loc}$ is usually related to the average macroscopic field, $E$, as follows:

$$E_{loc} = \beta E = \beta \frac{U}{d} \quad (2)$$

, where $\beta$ is the field enhancement factor, $U$ is the applied voltage and $d$ is the cathode – anode distance. Figure 5b shows the respective FN plots. The field enhancement factors are determined by fitting the linear part of each FN curve at low voltages, following Equation (1) and using the WF values obtained from UPS measurements. The corresponding dependence of $E_{to}$ and $\beta$ on the oxygen-to-carbon atomic percent incorporated into the rGO lattice, as determined by the respective XPS spectra in each case, is presented in Figure 6a. From Eq. (2) the local field required to initiate electron emission in each sample can be calculated. In Figure 6b we plot the dependence of the local field, $\beta E_{th}$, on the the oxygen to carbon atomic percent for the samples measured. A general trend shown by the these results is that the lower the carbon oxides content the lower the extraction field and the better the FE performance. This can be attributed to the increasing WF upon increasing the oxygen content, in agreement with literature [25]. Besides this, the UPS results presented above indicate that the density of states in the samples with lower oxygen content, i.e TR and NaOH(1) based, is higher than those based on KOH and NaOH(2), facilitating the electron transport at the substrate/cathode material interface. Moreover, concerning the role of Na on the performance of rGO/NaOH(1) and rGO/NaOH(2) cathodes, it should be emphasized that Na is present in the form of COONa compound and not in the form of an ion, therefore its contribution to the FE characteristics may be negligible [27]. Therefore we postulate that the field emission performance is not attributed to the Na presence, but only on the oxygen content.

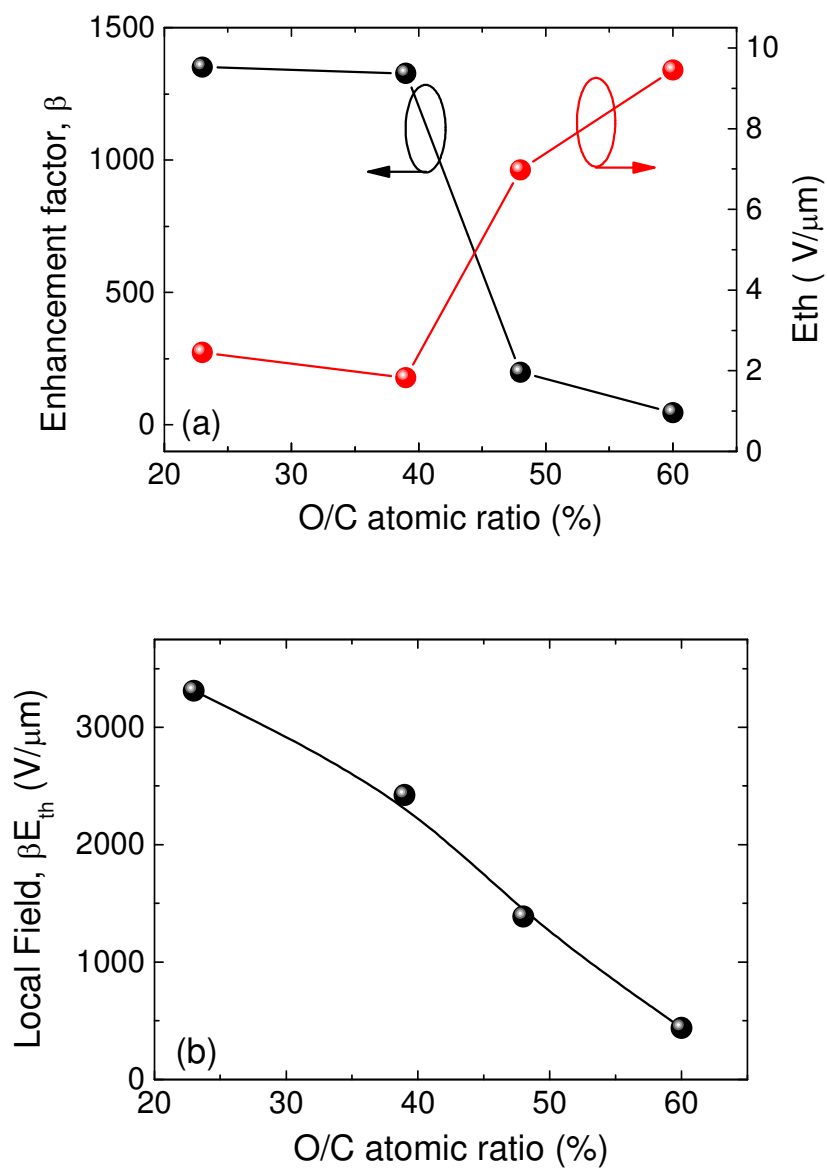

**Figure 6:** Dependence of the, $E_{to}$ and $β$ (a) as well as of the local field (b) on the oxygen atomic percent of carbon oxides attained following the application of different reduction methods.

The high field enhancement measured for the TR and NaOH(1) based rGO cathodes, i.e in the order of 1000, cannot be explained by the relatively smooth surface morphology of the rGO layers observed via SEM imaging (Figure 1). It should be noted that the $\beta$ values shown in Table 1 are determined using the WF of each respective cathode measured by UPS. Therefore, the geometric aspect ratio cannot account for the observed field enhancement and another emission mechanism should be explored. Yamaguchi et al. [28] reported that the emission from thermally reduced rGO cathodes occurs from atomically thin localized C-O-C chains of cyclic ether groups at the sheet edges. Such oxygen functional groups become predominant upon thermal reduction of GO lattice. In this context, the high field enhancement observed here may be attributed to atomically thin cyclic ether edges, that constitute the most stable form of oxygen in rGO, which are excellent linear sources of high-density electron beams [28].

Current saturation at high fields and thus the appearance of a knee point in F–N plots can be attributed to either resistive heating or high contact resistance effects. A similar saturation effect was observed in carbon nanotube films and attributed to adsorbents on the emitter tip [29] and a large voltage drop along the emitter and/or at the emitter/substrate interface [30][31].

Another important parameter that is crucial for device applications is the stability of the FE current over time. Figure 5c presents the evolution of the emission current density at a constant bias voltage of 1500 volts over a long period of continuous operation for the best performed rGO/TR cathode. The fluctuations of the emission current observed are commonly attributed to molecular adsorption on and/or ion bombardment of the emitting sites by residual gases, both of which are favored under high-vacuum conditions for graphitic based emitters [32]. However, resistive heating of the emitting apex have been experimentally observed and thought as the main reason of emission degradation [33].

## 4. CONCLUSIONS

In this work, we have performed a comparative study of the FE performance of rGO-based cathodes, which have been derived using rGO flakes produced by three different reduction methods. Spectroscopic analyses, such as XPS and UPS revealed that the various reduction methods lead to different oxygen functionalities and respective WFs giving rise to remarkable differences in the FE characteristics. Furhtermore, it is found that the oxygen content within the graphene lattice is the most critical parameter affecting the FE performance of rGO-based cathodes. The remarkable field enhancement observed and quantitatively evaluated via FN analysis revealed that the emission sites cannot be geometrically defined and should probably emanate from atomically thin regions.

## ACKNOWLEDGMENTS

This research has been co-financed by the European Union (European Social Fund – ESF) and Greek national funds through the Operational Program "Education and Lifelong Learning" of the National Strategic Reference Framework (NSRF) – Research Funding Program: Thales MIS 380389

# REFERENCES

1 H.Sugie, M.Tanemura, V.Fillip, K.Iwata, K. Takahashi, F.Okuyama, *Appl. Phys. Lett.*, 2001, **78**, 3578–2580.

2 G.Z.Yue, Q. Qiu, B.Gao, Y. Cheng, J.Zhang, H.Shimoda, S. Chang, J. P. Lu, O. Zhou, *Appl. Phys. Lett.*, 2002, **81**, 355–357.

3 L. F.Velasquez-Garcia, A. I.Akinwande, M. J.Martinez-Sanchez, *J. Microelectromech. Syst.*, 2006, **15**, 1272–1280.

4 C. Berger, Z. Song, X. Li, X. Wu, N. Brown, C. Naud, D. Mayou, T. Li, J. Hass, A. N. Marchenkov, E. H. Conrad, P. N. First, W. A. de Heer, *Science* 2006, **312**, 1191-1196

5 G. Eda, H. E. Unalan, N. Rupesinghe, G. A. J. Amaratunga, M. Chhowalla, *Appl. Phys. Lett.*, 2008, **93**, 233502.

6 E. Stratakis, G. Eda, H. Yamaguchi, E. Kymakis, C. Fotakis, M.Chhowalla, *Nanoscale*, 2012, **4**, 3069-3074.

7 I. Lahiri, V. P. Verma, W. Choi, *Carbon*, 2011, **49**, 1614-1619.

8 K. S. Novoselov, D. Jiang, F. Schedin, T. J. Booth, V. V. Khotkevich, S. V. Morozov, A. K. Geim, *Proc. Natl. Acad. Sci. USA*, 2005, **102**, 10451-3.

9 X. Li, W. Cai, J. An, S. Kim, J. Nah, D. Yang, R. Piner, A. Velamakanni, I. Jung, E. Tutuc, S.K. Banerjee, L. Colombo, R.S. Ruoff, *Science*, 2009, **324,** 1312-4.

10 Y. Hernandez, V.Nicolosi, M.Lotya, F.M. Blighe, Z. Sun, S. De, I.T. McGovern, B. Holland, M. Byrne, Y.K. Gun'Ko, J.J. Boland, P. Niraj, G. Duesberg, S. Krishnamurthy, R. Goodhue, J. Hutchison, V. Scardaci, A.C. Ferrari, J.N. Coleman, *Nat. Nanotechnol.*, 2008, **3,** 563-8.

11 X. An, F.Liu, Y.J.Jung, S.J. Kar, *J. Phys. Chem. C*, 2012, **116**, 16412-16420.

12 D.A. Dikin, S. Stankovich, E.J Zimney, R.D Piner, H.Geoffrey, B.Dommett, G.Evmenenko, S.T .Nguyen, R.S.Ruoff, *Nature*, 2007, **448**, 457-460.

13 S.Wakeland, R.Martinez, J.K. Grey, C.C.Luhrs, *Carbon*, 2010, **48**, 3463-3470.

14 P. Sungjin; A. Jinho, J. R. Potts, A.Velamakanni, S. Murali, R.S. Ruoff, *Carbon*, 2011,**49**, 3019-3023.

15 O.Akhavan, *Carbon,* **2010**, 48, 509-519.

16 Y. Guo, B. Wu, H. Liu, Y. Ma, Y. Yang, J., G. Yu, Y. Liu, *Advanced Materials*, 2011, **23**, 4626-4630.

17 Z. S. Wu, S. Pei, W. Ren, D. Tang, L. Gao, B. Liu, F. Li, C. Liu, H. M. Cheng, *Adv. Mater.*, 2009, **21**, 1756–1760.

18 N.I. Kovtyukhova, J. Olivier, B.R. Martin, T.E. Mallouk, S.A. Chizhik, E.V. Buzaneva, A.D. Gorchinskiy, *Chem. Mater.*, 1999, **11**, 771.

19 G. Viskadouros, M.M. Stylianakis, E. Kymakis, E. Stratakis, *ACS Appl. Mater. Interfaces*, 2013, *6*, 388.

20 X. Fan, W. Peng, Y. Li, X. Li, S. Wang, G. Zhang, F. Zhang, *Adv. Mater.,* 2008, *20*, 4490

21 Y. Jin, M. Jia, M. Zhang, Q. Wen, *Appl. Surf. Sci.,* 2013, *264*, 787.

22 E. Stratakis, R. Giorgi, M. Barberoglou, Th. Dikonimos, E. Salernitano, N. Lisi, E. Kymakis, *Appl. Phys. Lett.,* 2010, **96 (4)**, art. no. 043110.

23 A. Nikolakopoulou, D. Tasis, L. Sygellou, V. Dracopoulos, C. Galiotis, P. Lianos, *Electrochim. Acta,* 2013, **111**, 698.

24 P. Sungjin, A. Jinho, R. D. Piner, I. Jung, D. Yang, A.Velamakanni, S. T. Nguyen, and R. S. Ruoff, *Chem. Mater.,* 2008, **20 (21)**, 6592.

25 M. A. Munoz-Marquez, M.Zarrabeitia, E. Castillo-Martínez, A.Eguía-Barrio, T.Rojo, M. Casas-Cabanas, *ACS Appl. Mater. Interfaces*, 2015, DOI: 10.1021/acsami.5b01375

26 R. Garg, N.K.Dutta, N.R.Choudhury, *J. Nanomater.*,2014, **4**, 267-300.

27 Jen-Hsien Huang, Jheng-Hao Fang, Chung-Chun Liu, Chih-Wei Chu, *ACS Nano,* 2014, **8**, 6262.

28 H. Yamaguchi, K. Murakami, G. Eda, T. Fujita, P. Guan, W. Wang, C. Gong, J. Boisse, S. Miller, M. Acik, K. Cho, Y. J. Chabal, M. Chen, F. Wakaya, M. Takai, M. Chhowalla, *ACS Nano* 2011, **5**, 4945.

29 A. T. H. Chuang, J. Robertson, B. O. Boskovic, K. K. K.Koziol, *Appl. Phys. Lett.*, 2007, **90**, 123107.

30 E. Minoux, O. Groening, K. B. K. Teo, S. H. Dalal, L. Gangloff, J. P. Schnell, L. Hudanski, I.Y. Y. Bu, P.Vincent, P. Legagneux, G. A. J. Amaratunga, W. I. Milne, *Nano Lett.,* 2005, **5**, 2135.

31 W. J. Zhao, W. Rochanachivapar, M. J. Takai, *J. Vac. Sci. Technol. B*, 2004, **22**,1315

32 Q. M. Min, T. Feng, H. Ding, L. Lin, H. Li, Y. Chen, Z. Sun, *Nanotechnology*, 2009, **20**, 425702.

33 V. K. Kayastha, B. Ulmen, Y. K. Yap, *Nanotechnology*, 2007, **18**, 035206.

## Graphical Abstract

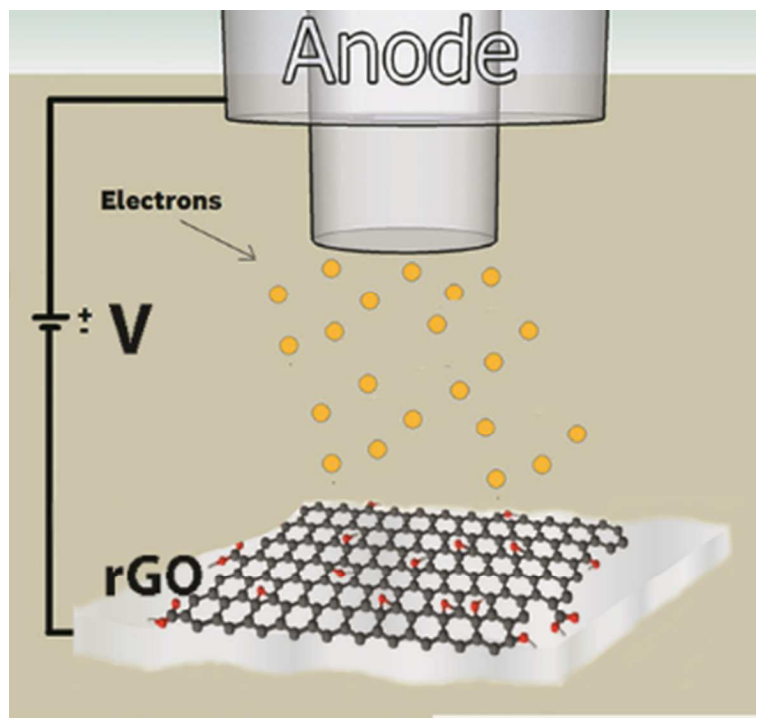

The effect of the reduction process and oxygen-contained functional groups on the field emission performance of reduced graphene oxide cathodes.